\documentclass{article}

\usepackage{arxiv}

\usepackage[utf8]{inputenc} 
\usepackage[T1]{fontenc}    
\usepackage{hyperref}       
\usepackage{url}            
\usepackage{booktabs}       
\usepackage{amsfonts}       
\usepackage{nicefrac}       
\usepackage{microtype}      
\usepackage{lipsum}
\usepackage{graphicx}
\usepackage{amsmath}
\usepackage[]{natbib}
\usepackage{float}

\title{Characterizing Evaporation Ducts Within the Marine Atmospheric Boundary Layer Using Artificial Neural Networks}

\author{
  Hilarie Sit\\
  Civil and Environmental Engineering\\
  Cornell University\\
  Ithaca, NY 14853 \\
  \texttt{hs764@cornell.edu} \\
   \And
  Christopher J. Earls \\
  Civil and Environmental Engineering\\
  Center for Applied Mathematics\\
  Cornell University\\
  Ithaca, NY 14853 \\
}

\begin{document}
\maketitle

\begin{abstract}
We apply a multilayer perceptron machine learning (ML) regression approach to infer electromagnetic (EM) duct heights within the marine atmospheric boundary layer (MABL) using sparsely sampled EM propagation data obtained within a bistatic context. This paper explains the rationale behind the selection of the ML network architecture, along with other model hyperparameters, in an effort to demystify the process of arriving at a useful ML model. The resulting speed of our ML predictions of EM duct heights, using sparse data measurements within MABL, indicates the suitability of the proposed method for real-time applications.
\end{abstract}


\section{Introduction}
The marine atmospheric boundary layer (MABL) is a dynamic region in the lower troposphere that directly interacts with the ocean via turbulent transport of heat, moisture and momentum \citep{Sikora}. Atmospheric conditions within the MABL, including its high temperature, pressure, and humidity gradients, influence the refractive index; thus affecting electromagnetic (EM) wave propagation characteristics. Rapid decrease in the refractive index with altitude can cause EM waves to bend backwards toward the ocean surface \citep{Skolnik}. This type of behavior is called atmospheric ducting, and results in the trapping of EM waves near the ocean free surface. Atmospheric ducting can influence radar performance by creating unexpected holes in the radar coverage, inaccurate altitude predictions, and false targets from clutter returns \citep{Skolnik}. Thus, real-time identification and characterization of EM ducting within the MABL is important when assessing radar performance as shown in Figure 1.

\begin{figure}[h]
\centering
\includegraphics[width=0.5\linewidth]{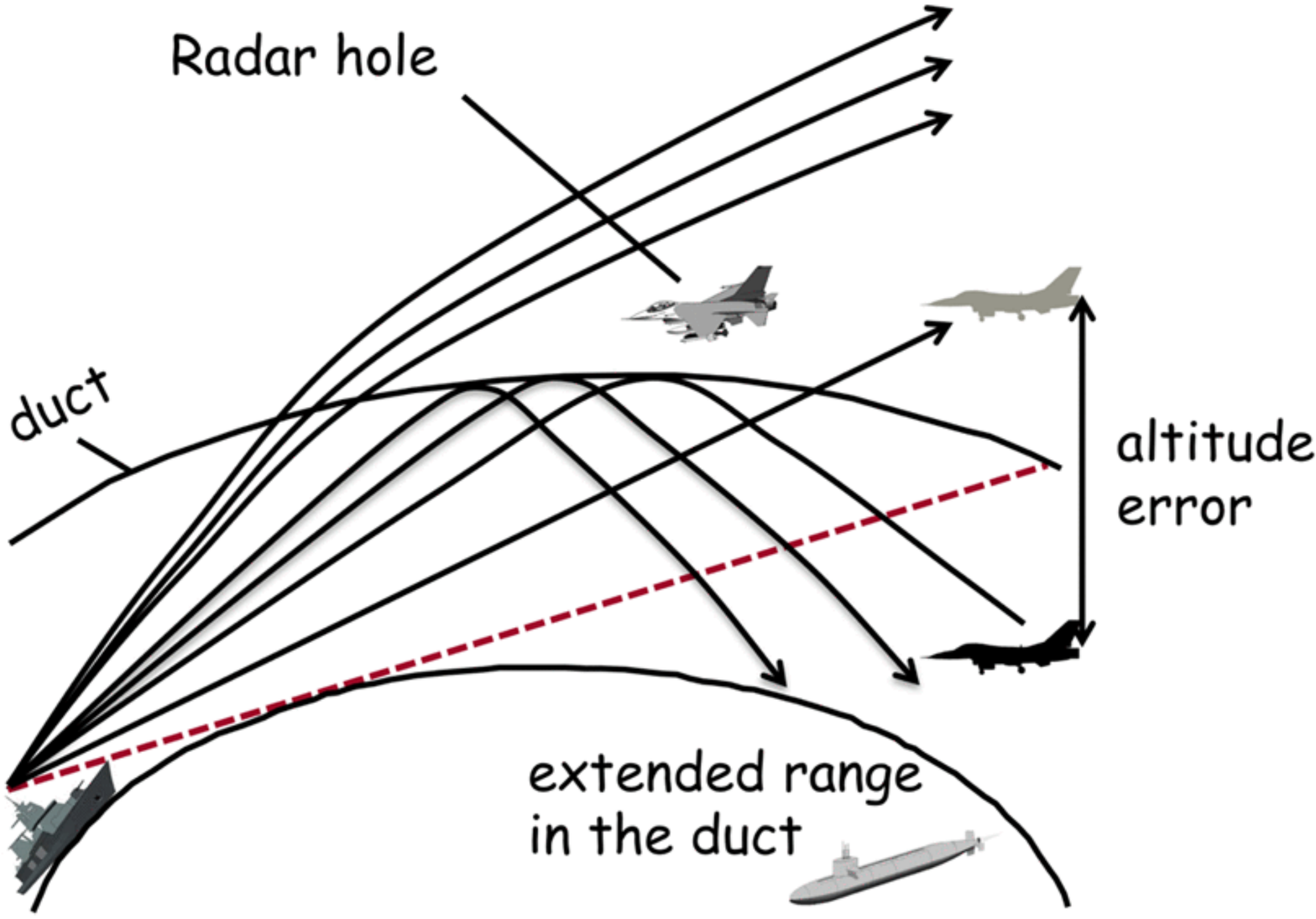}
\caption{Effect of duct on radar operations \citep{Fountoulakis}}
\label{figone}
\end{figure}

The refractive index can be approximated by directly measuring atmospheric conditions within the MABL: \cite{Bean} describe a relationship between the refractive index profile and atmospheric temperature, pressure, and humidity. Sensitivity of the various atmospheric and refractivity profile parameters in modeling EM propagation is detailed in \cite{Lentini}. Direct measurement of atmospheric parameters can be carried out at discrete altitudes by using radiosondes or rocketsondes, but such measurements are impractical when estimating MABL refractivity because data collection is sparse and implementation can be expensive \citep{Bean}. 

Recently, literature on the MABL EM duct characterization problem has focused on identifying and improving optimization within the context of the inverse problem estimating refractivity from clutter (RFC) returns. RFC uses measured propagation loss from radar clutter returns to estimate the refractivity profile, which is then used to characterize ducts. \cite{Rogers} show that duct heights can be inferred from a slope fitted to backscattered power from sea surface clutter, by performing a nonlinear least squares inversion. One common approach to solving the RFC inverse problem employs multiple calls of a suitable forward solver to predict EM propagation, but this can be computationally expensive. Efforts to make RFC inversion more efficient can also be found in the literature, including matched‐field processing approaches \citep{Gerstoft}, Markov-chain Monte Carlo \citep{Yardim}, and Markov state-space models \citep{Vasudevan}. A detailed review of RFC is available in \cite{Karimian}. 

In a different approach, \cite{Fountoulakis} develop a simplified physics model by approximating a blurring operator (so as to approximate the effects of the MABL) with modes from proper orthogonal decomposition (POD) from field observations that are organized within an offline library. Manifold interpolation is performed on the sparse set of POD modes to construct a real time, online estimate for POD modes, which are needed within an optimization algorithm to infer duct parameters. Within the context of this method, the coverage is either sampled along a sinusoidal flight path or a linear path (consistent with a rocketsonde-receiver sampling system, RRSS). RRSS consists of a receiver attached to a solid rocket flown at a constant altitude; thus effecting coverage sampling in a reduced amount of time (as compared to a UAV flying the sinusoidal path). In this paper, as part of our machine learning MABL EM duct prediction, we will use two bistatic (\textit{i.e.} separate transmitter and receiver locations) EM sampling schemes corresponding to: 1) a single moving receiver exploiting the practicality of RRSS, and 2) an array of stationary receivers mounted on a fixed tower (Figure 2).

\begin{figure}[h]
\centering
\includegraphics[width=0.9\linewidth]{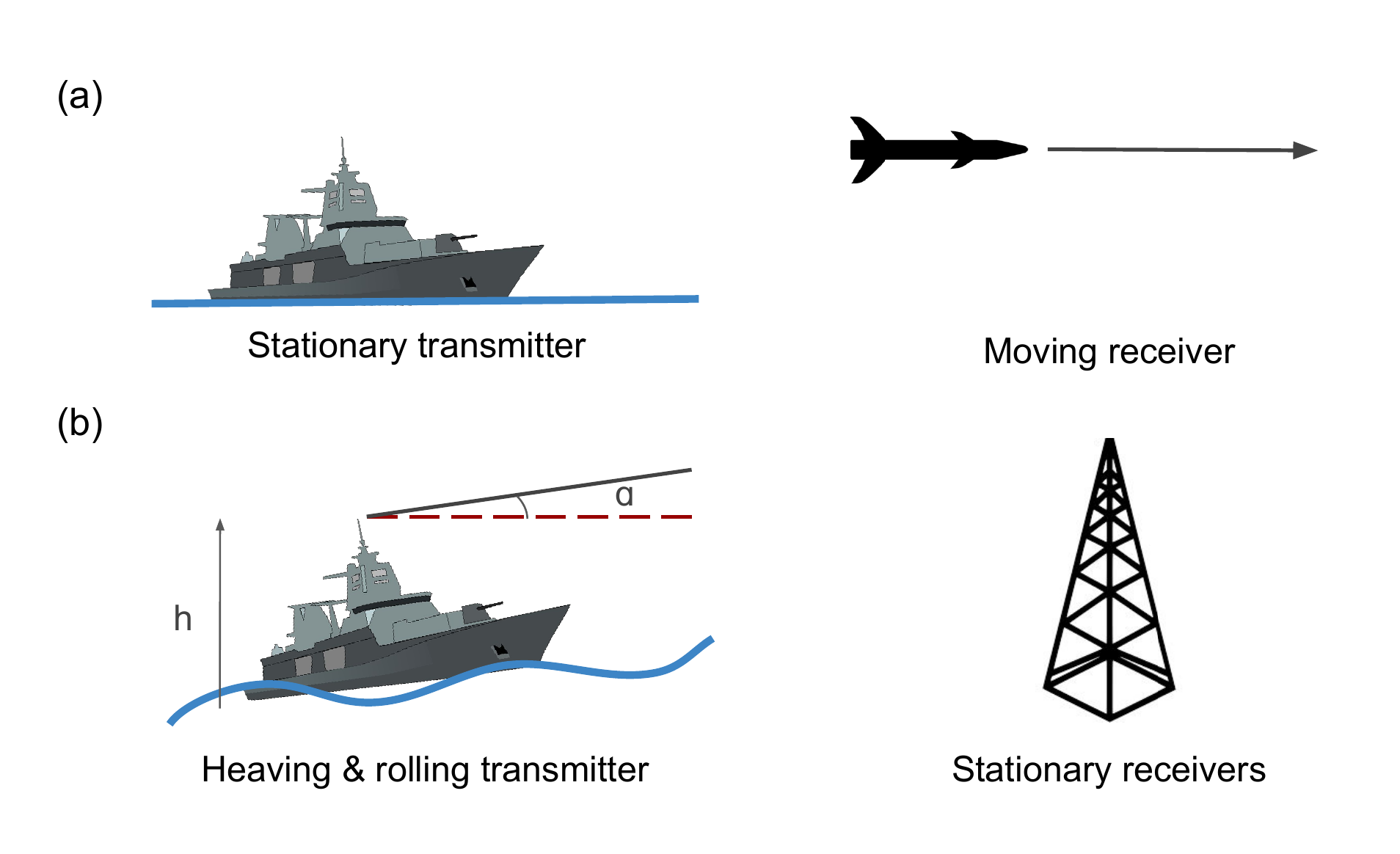}
\caption{Two bistatic radar sampling schemes: (a) moving receiver and static transmitter height and angle, and (b) stationary recievers on tower and varying transmitter heights and angles}
\label{figone}
\end{figure}

Machine learning algorithms have previously been employed within RFC inversion literature. \cite{Douvenot} showed that least squares support vector machines (LS-SVM) can speed up inversion, although sacrificing some accuracy, by utilizing a pre-generated database of propagation losses and respective parameter values. Within the RFC literature, variants of artificial neural networks (ANNs), primarily radial basis function neural networks, were considered but discounted in favor of the convex property of LS-SVM. A comparison of both approaches can be found in \cite{Yang}. With the recent surge in popularity of machine learning algorithms and deep learning models, enhancements to ANN architecture and optimization has led to increased model flexibility, stability, and efficiency. Recent literature has seen a revival in the use of ANNs for the duct characterization problem within an RFC context. \cite{Tepecik} introduce a hybrid model using parameter estimations from neural networks as initializers in genetic algorithms to refine the inversion. \cite{Guo} use deep neural networks to estimate refractivity parameters from calculated sea clutter power.

In the present paper, we propose using supervised machine learning with ANNs to directly estimate the function that maps propagation factors into duct height. ANNs are universal approximators with the ability to represent complex nonlinear functions using a few parameters \citep{Shalev-Shwartz}. Trained models can evaluate novel inputs and provide fast predictions; thus making ANNs potential candidates for real-time duct height predictions. We train and evaluate separate ANNs to estimate duct height using the aforementioned two bistatic sampling schemes (not RFC). Propagation data needed for training the ANNs are obtained using a fast split-step parabolic equation (SSPE) solver, as described in the sequel. 

This paper is outlined as follows. We begin by describing the physical problem and its domain. We subsequently introduce two bistatic sampling schemes along with the SSPE forward solver used in the numerical simulations needed to generate training and test data. We then provide a background on supervised machine learning with multilayer perceptrons and describe our specific machine learning regressive model. Finally, we report our results and provide a discussion of limitations in the model selection process.

\section{Forward Model}
\subsection{Evaporation Ducts} Evaporation ducts are the most prevalent duct type within the MABL \citep{Babin}. These shallow ducts form as a result of trapping gradients that change the refractivity profile over the ocean surface \citep{Paulus}. The refractive index, $n$, is the ratio of EM propagation velocity in free space to that within a given medium. To better reflect the small changes in the refractive index that characterize the MABL, we employ a quantity known as refractivity, $N = (n-1)10^6$ \citep{Skolnik}. The modified refractivity profile for evaporation ducts can be modeled as a log-linear function that is assumed to be homogenous in the range direction:

\begin{equation}
M(z) = M_0 + c \left(z - z_d ln\left(\frac{z+z_0}{z_0} \right) \right) 
\end{equation}

\noindent where $M_0$ is base refractivity, $c$ is critical potential refractivity, $z_0$ is aerodynamic surface roughness of the ocean, $z_d$ is the duct height, and $z$ is the altitude \citep{Fountoulakis}. Evaporation duct heights are commonly observed to be between $0m$ and $40m$ in altitude \citep{Paulus}. With the following parameters as constants: $M_0 = 428.89 M\textrm{-units}$, $c = 0.13 M\textrm{-units}/m$, and $z_0 = 0.00015m$, we are interested in the inversion for evaporation duct height, $z_d$.

\subsection{Problem Domain and Description}
In the present work, a horizontally-polarized Gaussian antenna pattern with radar signal frequency of 9.3 GHz, corresponding to the X-band, is transmitted through a rectangular 2D problem domain with a maximum altitude of $113m$ and range of $50km$. The associated SSPE problem domain is discretized using grid spacings of $0.1m$ and $40m$, in altitude and range, respectively. Propagation factors within the problem domain, collectively called the coverage diagram, are the observables, or feature vector, within the context of this work. In radar calculations, \textit{propagation factors} account for environmental effects and surface roughness and can be obtained by scaling the electric field magnitude with that observed in free space \citep{Ryan}.

Rather than using the entire coverage diagram to train our ANNs (as it is impractical to measure such large areas), propagation factors are sampled so as to be consistent with two bistatic radar sampling schemes mentioned earlier (Figure 2). A bistatic radar system uses different antennas for transmitting and receiving signals \citep{Skolnik}. We alternatively consider a moving receiver, as well as an array of stationary receivers mounted on a tower, to examine how variations in data collection may potentially influence ANN duct height estimates. For the moving receiver, the transmitter antenna is stationary with an antenna angle of $\alpha = 0^\circ$ (with respect to the horizontal) and antenna height of $h = 10m$. Propagation factors are sampled at 250 points along a horizontal path over the range, 5-15 km, positioned at 21m above the ocean free surface.

In the case of the array of stationary receivers, we consider two different configurations for the transmitter antenna, deterministic rocking and stochastic rocking, to enable the investigation of any influence from patterning within the resulting data sets. For deterministic rocking, six observations are collected using deterministic combinations of transmitter antenna angles $\alpha = -0.5^\circ, \ 0^\circ, \ 0.5^\circ$  and heights $h = 20m,\ 30m$ to simulate a transmitter antenna on a rocking ship. In this arbitrary setup, the expected heave of the ship is 10m with rolls of $\pm 0.5^\circ$. By altering the antenna height and angle in steady increments, this configuration leads to a defined pattern in the dataset, which is in contrast to the stochastic rocking configuration. Stochastic rocking is considered by collecting five observations from uniformly distributed antenna angles $\alpha \sim U[-0.5^\circ, 0.5^\circ]$ and heights $h \sim U[20m, 30m]$. Propagation factors are sampled using 30 evenly spaced receiver antennae distributed along on a vertical mast, 1-30m in the altitude direction, positioned 50km down range from the rocking transmitter. Observations are concatenated to form observation arrays of lengths 180 and 150, for the deterministic and stochastic rocking, respectively. Please refer to Figure 3 for a depiction of the two sampling paths. In order to form surrogate experimental observations, the modeled datasets are also contaminated with Gaussian noise of $N (0, \sigma); \ \sigma = 0.1||x||_{\infty}$, where standard deviation is 0.1 times the absolute value of the largest propagation factor in an observation array, to simulate severe electronic sensor noise \citep{Fountoulakis}. Overall, we assess six cases described in Table 1.

\begin{figure}[h]
\centering
\includegraphics[width=0.8\linewidth]{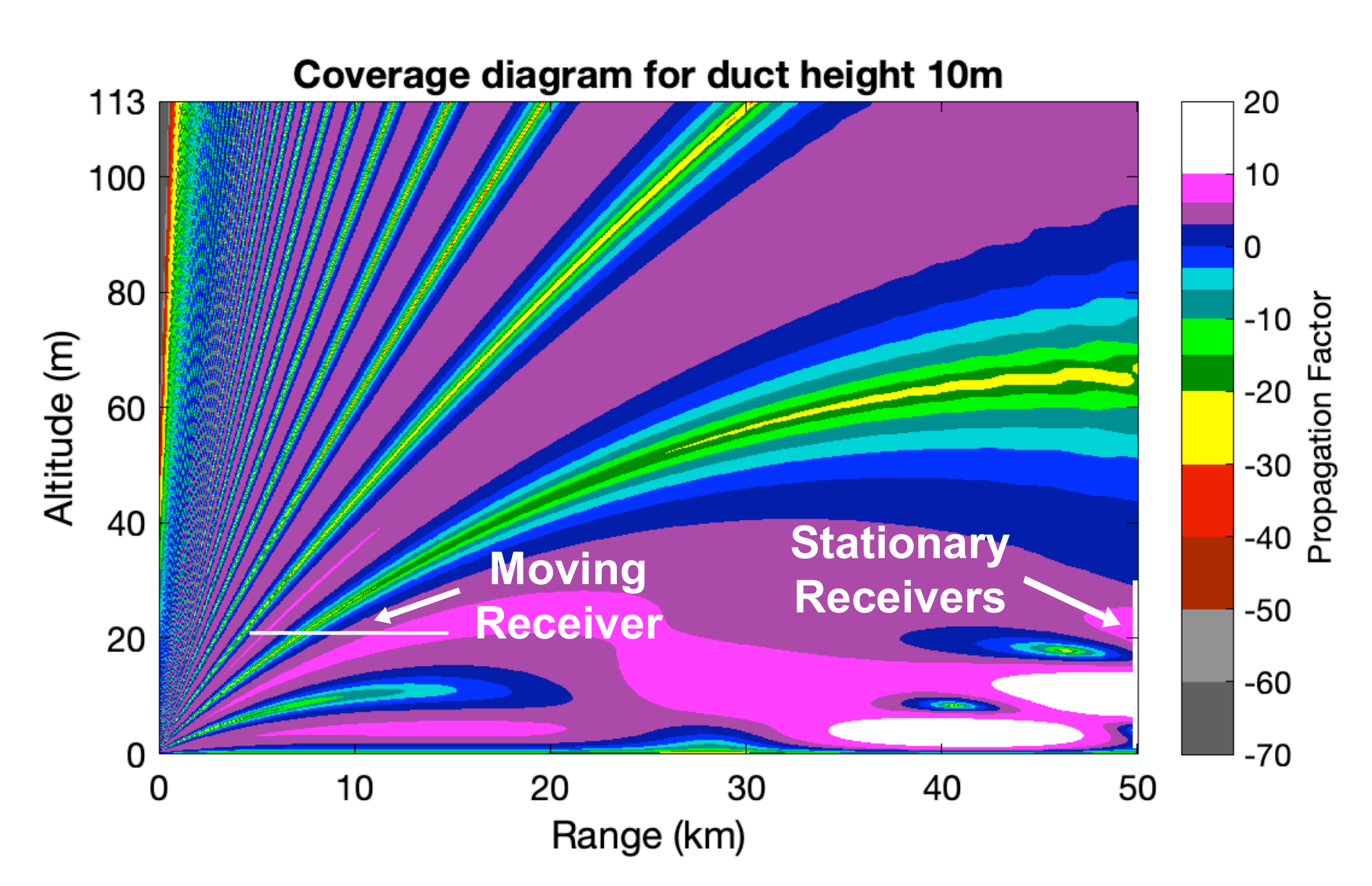}
\caption{Sampling paths on representative coverage diagram}
\label{figone}
\end{figure}

\begin{table}[h]
\centering
\caption{Description of the Cases Explored in this Paper}
\centering 
\begin{tabular}{l c}
\hline
Case 1: & Stationary transmitter, moving receiver (RRSS) \\
Case 2: & Deterministically rocking transmitter, stationary receivers \\
Case 3: & Stochastically rocking transmitter, stationary receivers \\
Case 4: & Stationary transmitter, moving receiver with noise (RRSS) \\
Case 5: & Deterministically rocking transmitter, stationary receivers with noise\\
Case 6: & Stochastically rocking transmitter, stationary receivers with noise \\
\hline
\end{tabular}
\end{table}

\subsection{SSPE Solution} EM propagation within the MABL can be modeled using the parabolic equation approximation to the Helmholtz wave equation. \citep{Ozgun} developed PETOOL, a MATLAB\textsuperscript{\tiny\textregistered}-based software, for analyzing EM propagation over variable terrain using the split step parabolic equation (SSPE) method. SSPE solves the wide angle parabolic equation as an initial value problem. The initial condition is specified by injecting EM fields into the domain at the transmitter's location, in order that a series of Fourier transformations can be performed to propagate the fields down range. PETOOL approximates the Sommerfeld radiation condition at the top of the problem domain by extending the domain's altitude and applying a Hanning window to remove the non-physical reflections that would ordinarily result from truncating our domain altitude at $113m$. The lower domain boundary is the ocean free surface, which is assumed to be smooth. To enforce continuity of the tangential components of the electric and magnetic fields at this boundary, the ocean surface is assumed to be a finite conductor with a homogeneous dielectric constant that can be calculated by using the semi-empirical Debye expression \citep{Ryan}:

\begin{equation}
\epsilon(\omega) = \epsilon_{ir} + \frac{\epsilon_{0} - \epsilon_{ir}}{1-i\omega\tau} + \frac{i\sigma}{\omega\epsilon_{0}}
\end{equation}

\noindent where the far-infrared dielectric constant of water, $\epsilon_{ir}$, is $4.9$, and the relaxation time, $\tau$, ionic conductivity, $\sigma$, and static dielectric constant of sea water, $\epsilon_0$, are obtained using thermodynamic data consist with the South China Sea - 100\% humidity at ocean surface, surface temperature of 29.7 $^{\circ}C$, and ocean salinity of 35 ppt. We use code adapted from PETOOL that specifies this Leontovich surface impedance condition at the lower boundary \citep{Gilles}.

\section{Machine Learning Regressive Model}
\subsection{Background}
The goal of supervised machine learning is to learn an hypothesis $h: X \rightarrow Y$ that maps inputs to outputs by only sampling points from some underlying distribution $(x_i, y_i) \in X \times Y$ \citep{Shalev-Shwartz}. The machine learning (ML) task of performing MABL duct height inversion is a regression problem, because duct height is a continuous variable. The parameters of the hypothesis (\textit{i.e.} our ML regression network weights and biases) can be learned by minimizing a loss function on training pairs (\textit{i.e.} known sets of propagation samples at given duct heights). Mean Squared Error (MSE), which measures the average squared difference between the predicted value and true value, is typically minimized when training ML regression networks \citep{James}:

\begin{equation}
J = \frac{1}{n} \sum^n_{i=1} (\hat{y}_i - y_i)^2
\end{equation}

\noindent where $\hat{y}$ is the output from the model (\textit{i.e.} the predicted duct height from the ANN) and $y$ is the label (\textit{i.e.} the true duct height). We are interested in employing artificial neural networks, a class of machine learning algorithms, for this task, because of their universal approximation property and quick evaluation speed. 

\begin{figure}[h]
\centering
\includegraphics[width=0.8\linewidth]{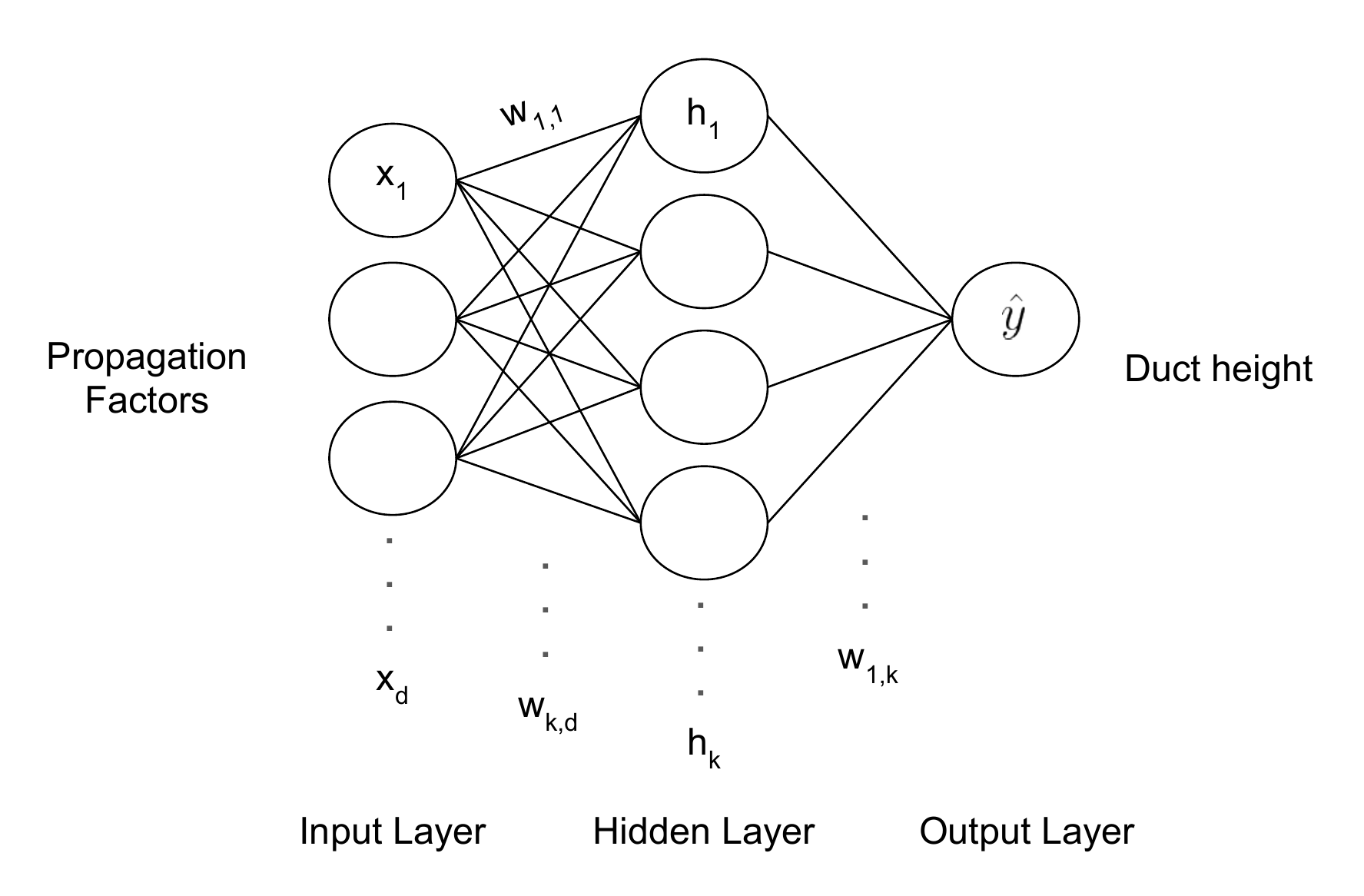}
\caption{MLP architecture}
\label{figone}
\end{figure}

Multilayer perceptrons (MLP) are feedforward artificial neural networks, consisting of an input layer, hidden layer(s), and an output layer. The input layer accepts a d-dimensional feature vector $x \in {\rm I\!R}^d$ (in our case, samples of propagation factors) where each dimension, $d$, corresponds to a neuron at a discrete spatial point (corresponding to a propagation factor measurement), such that each input neuron is connected to every neuron in the subsequent hidden layer by weighted edges within the network graph (Figure 4). The collection of network weights $w \in {\rm I\!R}^{k \times d}$ are summed together along with an offset term $b \in {\rm I\!R}^ k$; thus the value at a given hidden neuron can be written as:

\begin{equation}
h_k = \sum_d w_{k,d}x_d + b_k
\end{equation}

\noindent where $d$ is the input dimensionality and $k$ is the number of hidden neurons. The collection of weights and biases is referred to as the networks \textit{parameters}. Nonlinearity is added to the model by applying activation functions independently to the value at every hidden neuron. Common activation functions include sigmoid, hyperbolic tangent and rectified nonlinear unit (ReLU), with ReLU preferred due to its computational efficiency and reduction of the vanishing gradient problem \citep{Nair}:

\begin{equation}
ReLU(h) = max(0, h)
\end{equation}

\noindent The next hidden layer is connected to the previous hidden layer in the same manner as that of the input layer (\textit{i.e.} it is fully connected). The number of hidden layers and neurons in each layer are referred to as \textit{hyperparameters} (not to be confused with the network parameters in Eqn. 4). Hyperparameters are values that cannot be learned by the model and thus are carefully selected by the analyst prior to training, to achieve maximum predictive performance. In order to allow the MLP to predict a single continuous variable, the output layer must consist of a single neuron with no activation function: $\hat{y} \in {\rm I\!R}$. 

For MLPs, the MSE loss function is commonly minimized using a variant of gradient descent, an iterative, first-order gradient method. Higher-order optimization methods and quasi-Newton methods are not typically employed, as these methods can be slow to implement. In gradient descent, network parameters, $w$, are randomly initialized at the start of training and subsequently adjusted to improve the network's predictive power by taking an updating step in the opposite direction of the loss function's gradient with respect to the parameters, $\nabla J(w_{t})$ \citep{Shalev-Shwartz}:

\begin{equation}
  w_{t+1}  := w_{t}  - \alpha \nabla J(w_{t})
\end{equation}

\noindent where learning rate, $\alpha$, governs the step sizes of each iteration, and thus embodies a trade-off between convergence and speed. If $\alpha$ is too small, then the model takes too long to converge and conversely, if $\alpha$ is too large, the model can diverge. Gradients in neural networks can be efficiently calculated using the two-part \textit{backpropagation} algorithm. In the forward pass, inputs from the training set are fed into the neural network, and network activations are calculated and cached. These values are then retrieved during the backward pass to calculate the gradients with respect to every parameter using the chain rule from calculus, which describe the interrelationship between parameters from the back of the network to the front \citep{LeCun}. Gradients are obtained for each example in the training set, and subsequently averaged to get $\nabla J(w_{t})$ for the update in Eqn. 6.

The loss function landscape of neural networks is typically non-convex, and thus, a stochastic variant of gradient descent can help the model escape local minima and saddle points as well as prevent overfitting. In stochastic gradient descent (SGD), rather than calculating the gradient of the loss function using the entire training set, the direction at every iteration is determined by a randomly selected training example \citep{Shalev-Shwartz}. SGD is expected to behave similarly to gradient descent, but introduces noise, which could potentially limit convergence rate \citep{Bottou}. A solution to this problem is using the averaged gradient from a small number of training examples, referred to as a mini-batch, to update the parameters in order to reduce the variance of the gradient approximations \citep{Bottou2}. In a version of mini-batch gradient descent, the training set is randomly shuffled and mini-batches of training examples are sampled sequentially, without replacement, for the update. The training set is then reshuffled after every epoch, which is defined as the occurrence of every training example having been passed through, and subsequently updated, the model. 

\cite{Kingma} introduced Adam optimization, a stochastic optimization scheme that incorporates ``momentum'' with individual adaptive learning rates for parameters by using estimates of the gradient's first and second moments. The bias-corrected exponential moving average of the gradient, with decay parameter $\rho_1 = 0.9$, is used as the first moment estimate, and the bias-corrected exponential moving average of the gradient squared, with decay parameter $\rho_2 = 0.999$, is used as the second moment estimate. Updates are performed until convergence. We trained the MLPs with mini-batch Adam optimizer as the learning approach in the current work.

In supervised learning, the model is trained using a subset of the complete, labeled dataset, in order to measure model generalization when the network is applied to the excluded set. A statistical technique used to obtain a principled metric of model performance on unseen data is \textit{k-fold cross-validation}, where the dataset is split into \textit{k} folds, and each fold is used as the testing set for \textit{k} independent training sessions \citep{Shalev-Shwartz}. Common \textit{k} values for k-fold cross-validation are $k = 5$ and $k=10$, as these are shown to appropriately balance the bias-variance tradeoff and reduce computational time as compared to $k = n$, in the leave-one-out cross validation technique, where \textit{n} is the cardinality of the dataset \citep{James}. The losses from each fold are then averaged to achieve an aggregated cross-validation score, which can be used to compare different hyperparameter instances to define an appropriate MLP architecture.

\subsection{MLP Model Selection}
The specific MLP model design and architecture used in each of the six cases from Table 1 is influenced by the dimensionality of the dataset and involves the selection of appropriate hyperparameters, which include the number of hidden layers, number of neurons, number of epochs employed during training, initial learning rate, and batch size. Hyperparameter selection is a complex problem that unfortunately can require prior experience, as mentioned in \citep{Yang}. Barring prior experience, the analyst typically chooses hyperparameters via trial and error, so as to achieve a sufficient test loss or to satisfy some threshold cross validation score. We attempt to demystify this process by discussing important concepts, summarized in Figure 5, and the intuition behind our model selection process.

\begin{figure}[h]
\centering
\includegraphics[width=\linewidth]{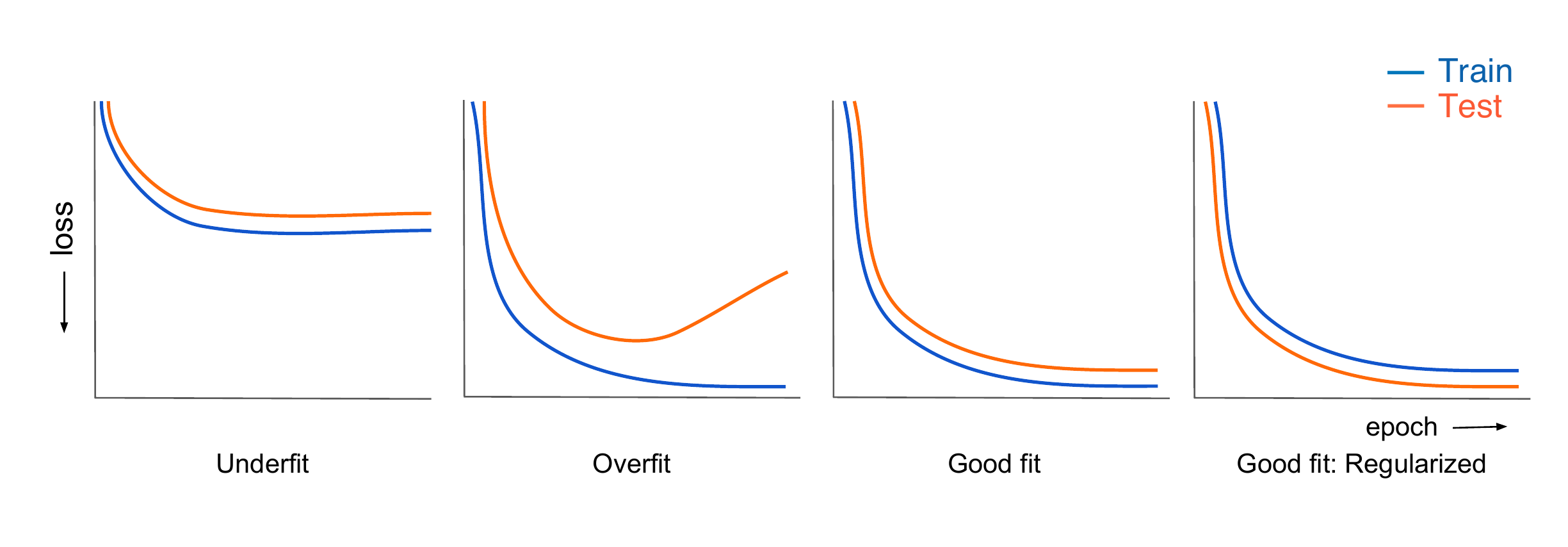}
\caption{Performance indicators for diagnosis of train and test loss behavior in training artificial neural networks}
\label{figone}
\end{figure}

A description of the four canonical network training behaviors from Figure 5 can be summarized as follows. 

\textbf{Underfitting.} Underfitting occurs when the model fails to sufficiently learn the underlying function approximation. When both training and test losses remain large, it is possible that the model has converged to a non-optimal local minimum or saddle point within the loss function output space. Underfitting can also be a result of having too few parameters to represent the pattern in the dataset. Increasing model complexity by increasing number of neurons as well as decreasing learning rate can help resolve underfitting. The model can also be underfitted if the selected number of epochs is too small, and the model has not yet converged since it has not seen the training examples enough times. 

\textbf{Overfitting.} Overfitting occurs when model performance on the training set far exceeds the performance on the test set \citep{Shalev-Shwartz}. This phenomenon is evident when test loss plateaus or increases as training loss decreases. For a small dataset, overfitting is a primary concern, because the number of parameters in the model may be too numerous for effective generalization to other, previously unseen, data. Methods to mitigate overfitting include stopping the training at an earlier epoch, decreasing the number of parameters within the model (\textit{e.g.} by either decreasing the neuronal cardinality within a layer and/or reducing the number of layers), and adding regularization.

\textbf{Regularized.} Regularization adds restrictions to the model, to address the overfitting problem, by forcing the network to learn a less complex representation by penalizing it for learning unwanted behaviors. A model with lower test loss than train loss is possible when using regularization techniques. Adding noise to the input is related to adding regularization to the weights of the ANN \citep{Bishop}.

\textbf{Spiky convergence.} Spiky convergence is common when using stochastic methods, such as using a mini-batch to update the parameters. Decreasing the learning rate or increasing the batch size can give smoother convergence. However, some stochasticity can be beneficial for escaping saddle points or local minima; thus care should be exercised when reducing its influence.

Trial and error methods based on intuition alone do not produce the best hyperparameters when designing a regressive MLP.  Rather, we use a more systematic approach by performing grid search in a restricted hyperparameter space. In this grid search, the number of hidden layers is set to one in all six cases and the number of epochs is chosen to be sufficiently large to allow for convergence for each case, as gauged by the loss function, Eqn. 3. For each of the noise-less cases (1-3), we explored 45 different model settings with the following hyperparameter combinations: hidden neuron cardinality within the single layer [180, 185, 190, 195, 200], learning rate [1e-4, 5e-4, 1e-3], and batch size [4, 8, 16]. For each of the noise-contaminated cases (4-6), we explored 60 different hyperparameter combinations: hidden neuron cardinality within the single layer [70, 75, 80, 85, 90, 95, 100, 105, 110, 115], learning rate [1e-4, 5e-4, 1e-3], and batch size [32, 64]. The model with the lowest 5-fold cross validation (CV) score from the training procedure detailed below is subsequently selected for each case as shown in Table 2. Other methods for hyperparameter optimization, besides grid search, include random search and bayesian optimization \citep{Bergstra}.

\begin{table}
\centering
\caption{Selected Models from Grid Search (assuming single hidden layer)} 
\begin{tabular}{l c c c c c c}
\hline
\ & Case 1 & Case 2 & Case 3 & Case 4 & Case 5 & Case 6 \\  [0.5ex] 
\hline
Input Dimensionality & 250 & 180 & 120 & 250 & 180 & 120 \\
Hidden Neurons & 200 & 180 & 195 & 100 & 70 & 95 \\
Epochs & 1000 & 2500 & 2500 & 400 & 500 & 500 \\
Learning Rate & 1e-3 & 1e-3 & 1e-3 & 1e-4 & 1e-3 & 5e-4\\
Batch Size & 16 & 4 & 4 & 32 & 32 & 32 \\
\hline
Loss mean / CV score $(m^2)$ & 1.34e-2 & 2.41e-1 & 2.75e-1 & 3.09e-1 & 2.78e-1 & 3.52e-1 \\
Loss s.t.d. $(m^2)$ & 8.42e-2 & 1.69e-1 & 1.47e-1 & 7.74e-2 & 9.89e-2 & 2.58e-1 \\
Accuracy mean $(\%)$ & 98.7 & 74.3 & 80.7 & 70.1 & 69.2 & 74.1 \\
Accuracy s.t.d. $(\%)$ & 2.67 & 19.1 & 5.23 & 3.46 & 15.0 & 5.56 \\
\hline
\end{tabular}
\end{table}

\subsection{MLP Network Training}
Datasets are constructed by sampling propagation factors corresponding to evaporation duct heights ranging from 2m to 40m, in half meter increments. The dataset is randomly split: 80\% for training and 20\% for testing. For cases 4-6, the inclusion of noise into the smaller dataset increases the difficulty of the problem. Overfitting is a common problem when training neural networks on a small dataset, as is the case in our problem. To address overfitting to the noise in cases 4-6, the training set is augmented by applying the Gaussian noise, $N (0, \sigma); \ \sigma = 0.1||x||_{\infty}$, to each training example instance (\textit{i.e.} at each considered duct height) 100 times. While the same noise treatment is applied to the observations in the test set, the test set is not augmented. Our multilayer perceptions were implemented using Tensorflow \citep{tensorflow}. The model weights are randomly initialized from a truncated gaussian distribution ($\mu = 0, \ \sigma = 0.01$) and biases are initialized as zeros. 

MSE (eq. 3) and accuracy are used as evaluation metrics when comparing MLP performance. Accuracy is normally a classification metric, but it can be informative to quantify instances of good predictions. In this paper, accuracy is defined as

\begin{equation}
A = \frac{1}{n}\sum^n_{i = 1} 1\!\rm l (|y-\hat{y}| < \delta) 
\end{equation}

\noindent where $1\!\rm l (\cdot) $ is the indicator function and $\delta = 0.5m$. This metric conservatively scores a difference of $0.5m$, or more, between the prediction and true duct height as incorrect. 

During training, test loss is calculated after every epoch and model parameters are saved for the best test loss. Both training and evaluation are performed using 12-cores of Intel Xeon E5 microprocessor having a clock speed of 2.7 GHz. Timing is calculated using the Python time.clock() method, which measures processor system time in seconds.

\subsection{Evaluation}
To determine the performance of the chosen model design and architecture, the entire dataset is once again split randomly into a training and test set, and the best models for each case from the 5-fold cross-validation hyperparameter exploration, is trained on the training set from scratch, and subsequently evaluated on the remaining test data. For each of the six cases, we report the MSE test loss, test accuracy, and absolute error of the worst prediction by the model. Training time is calculated by summing all forward pass and backpropagation calls with the training set, for the specified number of epochs in Table 2, and evaluation time is calculated as the average time to evaluate the entire test set. 

Training and test curves for both loss and accuracy during model evaluation are shown in Figures 6 and 7. A noticeable characteristic of the curves are the spikes that result from the stochasticity with using small batch sizes (bolded lines show a smoothed curve), which can help the model achieve better performance. Further inspection of the curves in relation to Figure 5 show that case 3 is overfitted: the validation loss is substantially higher than the training loss and the test accuracy is lower than the training accuracy. Table 3 also shows that the highest difference of the worst prediction and its true label occurs in case 6: the absolute error is $2m$, with the model predicting $38m$ for the true duct height of $40m$. The inclusion of dataset endpoint(s), $2m$ and $40m$, in the test set means that the ANN has to extrapolate beyond the range of the given training data, which can yield worse predictions. Variability in model selection and evaluation should be expected due to variability in parameter initializations and random dataset shuffling, splitting, and batching.

\subsection{Discussion}
A single hidden layer MLP, with its ability to provide quick and reasonable approximations of duct heights given propagation factors, is sufficient for the MABL duct height estimation task. We found that deep neural networks, which include two or more hidden layers, do not improve performance and can even hinder performance by contributing to the overfitting problem. Regularization techniques, to address the overfitting problem in neural networks exist, but these are not suitable in our problem. Very deep models can benefit from dropout \citep{Srivastava}, which involves randomly dropping out neurons to force the network to learn a more general representation with fewer parameters. Another technique is batch normalization \citep{Ioffe}, which involves normalizing inputs into a layer with the mean and standard deviation of batches, but this is not recommended for small batch sizes. It is pointed out that the limitations of small datasets within the current study, imposed on the models, are representative of problems in the physical world; where data are frequently difficult to obtain. This situation represents a challenge for many common machine learning techniques, when applied to problems of physics.

It is noted that the more formal nested cross-validation procedure is not performed within the current study due to the small dataset size. Instead, we evaluate model performance by re-splitting the dataset and re-training the model from scratch, but in general, re-training the model on the entire dataset and evaluating performance on a separate unseen set is better. Nested cross-validation addresses the potential problem of leaking validation set information into the model during the model selection process \citep{Cawley}. However, to do this, the dataset needs to be split into three sets for training, validating and testing, which our modeling context cannot afford (\textit{i.e.} we have far too few training pairs to accommodate this).

\begin{table}
\centering
\caption{Evaluation of Models} 
\begin{tabular}{l c c c c c c}
\hline
 \ & Case 1 & Case 2 & Case 3 & Case 4 & Case 5 & Case 6 \\  [0.5ex] 
\hline
Loss $(m^2)$ & 1.24e-2 & 4.81e-2 & 6.16e-1 & 3.22e-1 & 1.75e-1 & 4.57e-1\\
Accuracy $(\%)$ & 100 & 93.8 & 62.5 & 68.8 & 75.0 & 62.5 \\ 
Absolute Error $(m)$ & 0.305 & 0.568 & 1.82 & 1.26 & 0.829 & 2.00\\
\hline
Training time $(s)$ & 8.42 & 67.9 & 68.2 & 143 & 169 & 173\\
Evaluation time $(s)$ & 8.99e-4 & 8.83e-4 & 8.98e-4 & 9.01e-4 & 8.88e-4 & 8.93e-4\\
\hline
\end{tabular}
\end{table}

\newpage
\begin{figure}[H]
\centering
\includegraphics[width=\linewidth]{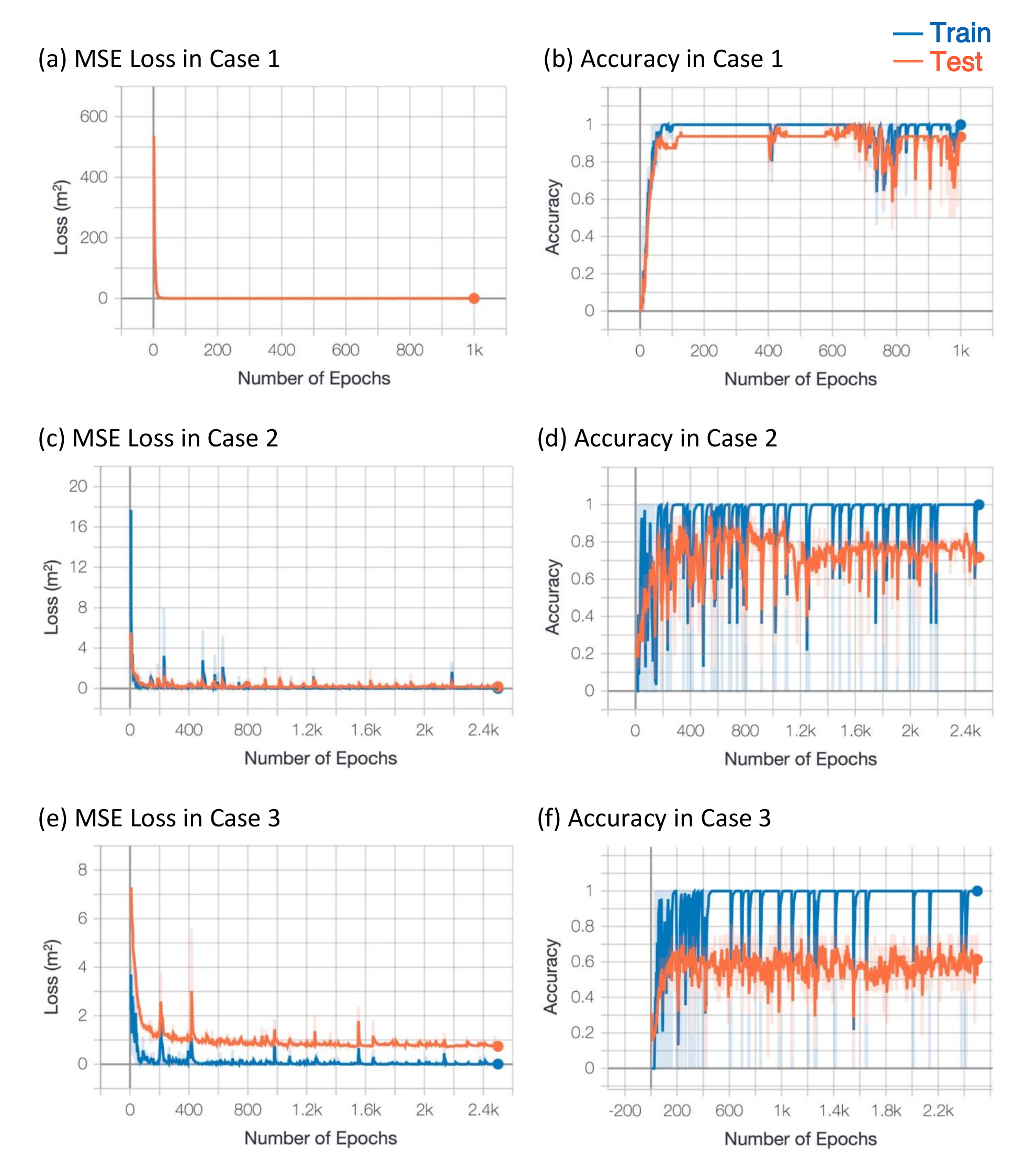}
\caption{MSE loss and accuracy for both the training set (blue) and test set (orange) during evaluation of the noise-free cases 1-3}
\label{figone}
\end{figure}

\begin{figure}[H]
\centering
\includegraphics[width=\linewidth]{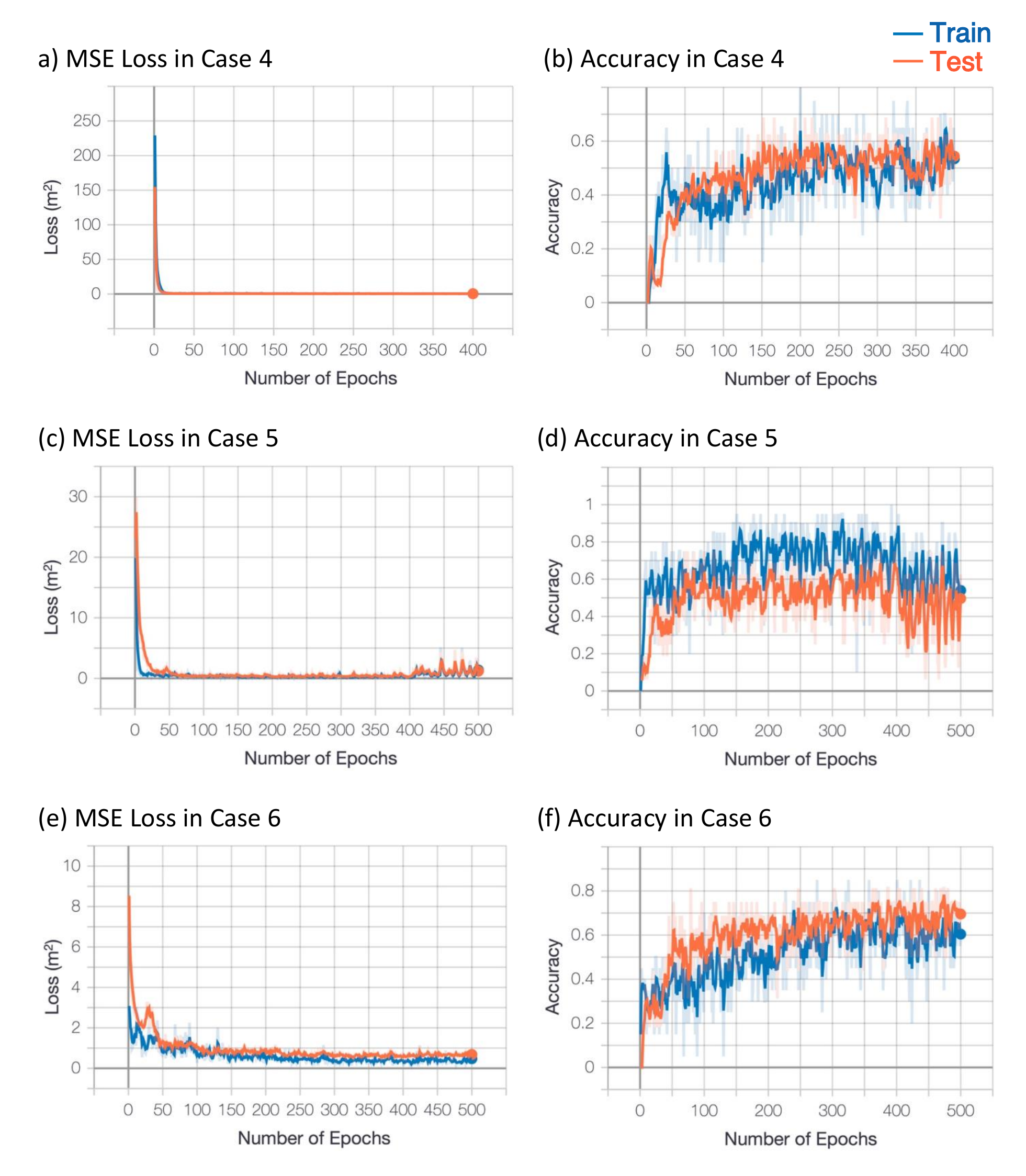}
\caption{MSE loss and accuracy for both the training set (blue) and test set (orange) during evaluation of the noise-contaminated cases 4-6}
\label{figone}
\end{figure}

\section{Conclusion}
Artificial neural networks show promising results in characterizing evaporation ducts within the MABL. We have shown that single hidden layer MLPs can sufficiently estimate duct heights from propagation factors under different bistatic sampling schemes, indicating the robustness of MLPs on these different data collection methods. It is noted that the results here were based on the use of surrogate data and involved certain idealizations (\textit{e.g.} smooth ocean surface, constant refractivity profile in range, etc.). However, it is expected that the proposed methods will generalize to even more realistic cases, but the training data must reflect these less idealized conditions.

A highlight of the present work is that MLPs are extremely efficient during evaluation, using much less than a millisecond of CPU time to predict duct height from data. As a result of the observed accuracy and precision along with evaluation speed, we conclude that ANNs are excellent candidates for real-time prediction of ducting conditions in the MABL.

\section{Acknowledgments}
Datasets and code for ANN selection and evaluation are available at https://github.com/nonlinearfun/ANN-selection-and-evaluation. The authors gratefully acknowledge ONR Division 331 \& Dr. Steve Russell for the financial support of this work through grant N00014-19-1-2095.

\bibliography{paperbib_arxiv}

\end{document}